# ALLEGRO Core Degradation Study using MELCOR 2.1

Petr Vácha

ÚJV Řež, Hlavní 130, 250 68
Husinec-Řež, Czech Republic, petr.vacha@ujv.cz

Ladislav Bělovský

ÚJV Řež, Hlavní 130, 250 68
Husinec-Řež, Czech Republic, ladislav.belovsky@ujv.cz

ALLEGRO is an experimental high-temperature gas (He)-cooled fast reactor (GFR) under development by the European Consortium "V4G4 Centre of Excellence" on basis of the concept ALLEGRO presented by CEA in 2009. Its main purpose is to demonstrate the viability of GFR in pilot scale and simultaneously serve as a test bed for GFR-related technologies, above all the high-temperature resistant refractory fuel in the conventional oxide fuel driver core. Severe accident studies of the concept with the first core (oxide fuel in stainless steel claddings) are an important part of safety analyses leading to improves in reactor design and safety. This paper extends our previous study using the MELCOR code. Major difference between the previous and this work is decrease of thermal power and power density to 50 MWth and 66,6 MWth/$m^3$ respectively. Analyses are focused on protected scenarios based on total station blackout individually aggravated by loss of primary coolant, water ingress into primary coolant or malfunction of check valves in the decay heat removal system. The results indicate the timing and extent of core degradation and provide valuable data for design of the core catcher in ALLEGRO.

## I. INTRODUCTION

Since no gas-cooled fast reactor (GFR) has ever been built, a small demonstration reactor is necessary on the road towards the full-scale GFR. A concept of this experimental fast reactor cooled with helium, called ALLEGRO, has been developed by an European consortium "V4G4 Centre of Excellence" (V4G4) since 2013. The consortium consists of nuclear research organizations of the Czech Republic (ÚJV Řež, a. s.), Hungary (MTA-EK Budapest), Poland (NCBJ Swierk) and Slovakia (VUJE a.s.) associated with the French Commissariat à l'énergie atomique et aux énergies alternatives (CEA). ALLEGRO is an important step on the way to the full scale Gas-cooled Fast Reactor (e.g. the GFR2400 concept by CEA[1]), one of the six concepts selected by the Generation IV International Forum, and one of the three fast reactors supported by the European Sustainable Nuclear Energy Technology Platform (SNETP).

Starting from the reference pre-conceptual design studied between 2001-2009 at CEA (ALLEGRO CEA 2009)[2], Fig 1, the project by V4G4 is exploring a new, lower target volumetric power density, reduced from the original 100 MWt/$m^3$, to be compatible with the safety limits and the design requirements. At the same time, the feasibility of a pin type LEU UOX start-up (first) core with stainless steel claddings as alternative to MOX fuel is being considered. This start-up core, to be used in the first period of the reactor operation, will include experimental positions dedicated to the refractory fuel development.

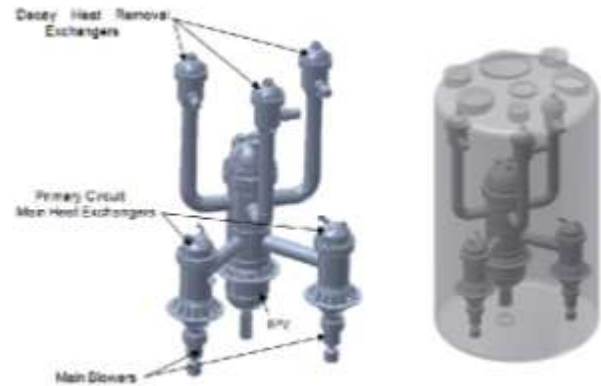

Fig 1. Primary circuit of ALLEGRO enclosed in a pressure boundary (guard vessel). Support structures and auxiliary systems are not shown.

Safety analyses including severe accident studies are part of the design process of ALLEGRO with the start-up core[3]. The weak point of this concept is the stainless steel widely used in the core. Its relatively low melting point (between 1300-1400 °C) makes the concept susceptible to overheating and, thus, also to severe fuel damage and to core meltdown.

The first severe accident study of the ALLEGRO CEA 2009 using MELCOR 2.1 has been published recently[4] and was aimed mainly at confirming the applicability of MELCOR to ALLEGRO. This study discusses the process of core degradation in ALLEGRO with the first MOX core and power limited to 50 MWt (~66 MWt/m$^3$). Computer code MELCOR version 2.1 was used to analyze protected scenarios characterized by total station blackout individually aggravated by loss of primary coolant, water ingress into primary coolant or malfunction of check valves in the decay heat removal system. The obtained experience enables us to start first qualitative & quantitative considerations about use of a core catcher in ALLEGRO.

## II. SEVERE ACCIDENT PHENOMENOLOGY IN ALLEGRO WITH THE FIRST CORE

The phenomenology of refractory core degradation is discussed in Ref. 5 and Ref. 6. The phenomenology of degradation of the first core with oxide fuel in stainless steel tubes is different from that of the refractory core. Large amount of molten materials can appear at temperatures well below 1500 ºC.

The most pronounced source of molten material during a SA in ALLEGRO is melting of the fuel assembly austenitic stainless steel claddings & wrapper tubes (15-15Ti /AIM1 above cca 1320 ºC) and materials of the core structural components (AISI-316: 1375-1400 ºC, AISI-304: 1400-1450 ºC).

### II.A. Material Interactions in SA conditions

Material interactions in SA conditions in ALLEGRO are governed either by exposition of materials to gas (nitrogen ingress from guard vessel during LOCAs, water steam due to water ingress from DHR loops) or by formation of eutectics in material couples in contact.

Water steam can cause high-temperature oxidation of the steel core internals such as fuel claddings and support plates, leading to degradation of their mechanical properties and possibly even to failure of their supporting function, as has been shown in Ref. 4.

Nitridation of steels, such as AISI-304, at high temperatures causes degradation of the mechanical properties, this process is, however, usually slower than the typical times of in-vessel phase of a severe accident[7] and contact of steel internals with hot nitrogen would not probably be the main reason of their mechanical failure. Immediate effect of hot nitrogen on AIM1 steel at high temperatures is currently unknown and experimental data are needed.

The most important among the chemical interactions leading to formation of eutectics is the one between the B$_4$C absorber (pellet or powder) and the austenitic control rod wrapper tube made of 15-15Ti /AIM1 stainless steel, where the eutectics forms at temperatures below 1300 °C (1250 ºC in case of AISI-304)[8]

### II.B. Core Degradation Mechanisms

Core degradation in ALLEGRO with the first core is driven mostly by the physical phenomena, due to very limited chemical interaction between the fuel pellets, cladding and, above all, coolant, which is an inert gas. It is characteristic with rapid propagation in the first (tens of) minutes after the disbalance in heat production and removal in the core occurs. It is followed by a slower evolution into the next phases of the severe accident. The slower onset of the later in-vessel phases is caused by relatively low total decay power produced in the degraded fuel, dropping to cca 700 kW after one hour from the reactor shutdown.

Melting and failure of the cladding can occur before the fuel reaches the melting temperature, due to very limited chemical interactions combined with very significant difference in melting temperatures of the cladding (cca 1600 K) and fuel (cca 3000 K). It could lead to compaction of the fuel pellets and other debris, causing its recriticality. In some extreme cases, this mechanism can lead even to prompt critical configurations.[9]

Fast excursions to temperatures over 1100 °C also causes the top of the two core support plates in ALLGERO, made of steel, to fail quite early, before significant amount of other core internals, such as the reflector sub assemblies, is melted. A scheme of the process of degradation with an estimation of the amount of melted materials relocating into the lower head is in Fig. 2. The estimations mentioned in the previous text are based on early explorative analyses of ALLEGRO core degradation carried out in MELCOR 1.8.6.

Several observations can be made, based on the information in the scheme. First, amount of oxides in the molten corium is relatively low, because the only oxide material present in the core is the MOX fuel. No additional oxidation should occur, except for the water ingress scenarios (atmosphere inside the GV is pure nitrogen).

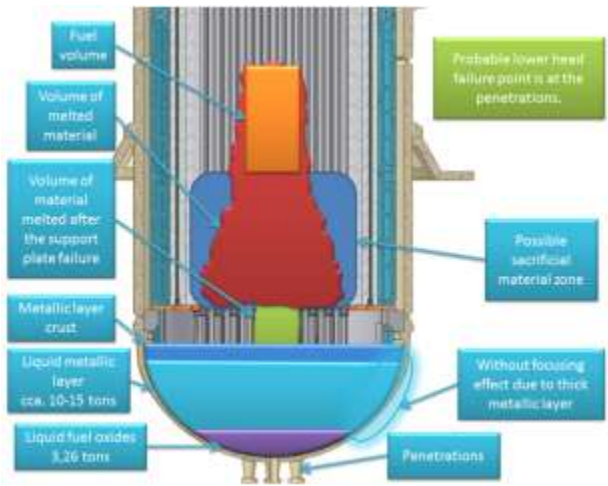

Fig. 2. Scheme of ALLEGRO core degradation.

Corium coming from molten ALLEGRO first core will contain a large amount of metallic phase, mainly steel from molten reactor internals,. This composition should suppress the focusing effect, and the most probable points of lower head failure are, therefore, located in the bottom penetrations used for the control rod driving mechanism (CRDM).

The lack of other oxide materials in the oxide phase contained in the lower head before its failure is yet another issue potentially leading to recriticality of the degraded core.

## III. MELCOR MODEL

A model of 75 MWth ALLEGRO based on CEA 2009 design was developed in UJV[4]. For the studies presented in this paper, the model has been adapted to include new changes in design. It means mainly reducing the nominal power to 50 MWth. while keeping the same inlet and outlet temperatures. It has been achieved by reducing the nominal coolant flow rate to 2/3 of the nominal value for the 75 MW reactor. Comparison of selected nominal values to those calculated in MELCOR is shown in TABLE 1.

TABLE 1. Nominal operation simulation

| Property | Target Value (75 MW) | Target Value (50 MW) | Calculated value |
|---|---|---|---|
| Nominal Power (MWth) | 75.0 | 50.0 | 50.0 |
| Core mass flow (kg/s) | 53.6 | 35.5 | 35.2 |
| Core inlet T (°C) | 260.0 | 260.0 | 258.5 |
| Core outlet T (°C) | 530.0 | 530.0 | 533.0 |
| Maximum cladding T (°C) | 610.0 | 610.0 | 602.6 |

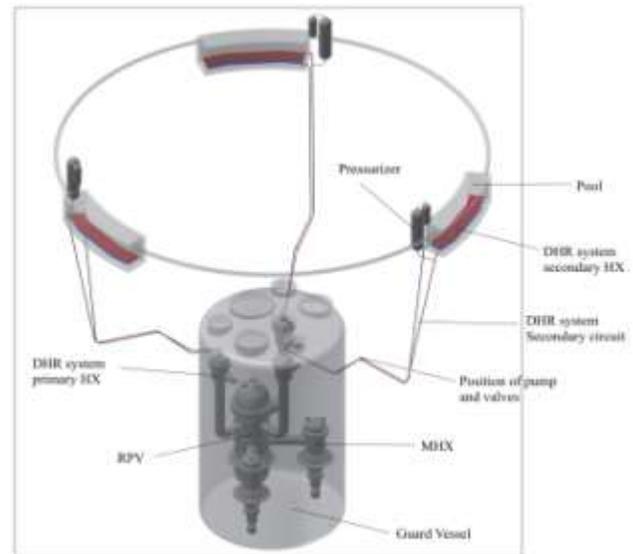

Fig. 3. Model of DHR system water circuits

Another significant modification of the MELCOR model is rearrangement of the secondary (water) circuit of the DHR system. In the original model it was assumed that the main part of the circuit is common for all the three loops and it splits just above the He/water heat exchangers to three separate loops. In the new model, there are three separated water circuits, interconnected by a duct on top of the final pools. A 3D CAD model of the circuits is shown in Fig. 3.

Overall core geometry model remains the same as for the 75 MWth version - BWR, which represents the actual core of ALLEGRO the best. The only problem with using this global core model is that it automatically puts non-supporting structures such as reflector SAs and control rods in the bypass control volumes, which is not the case in real ALLEGRO. It has been solved for the steel non-supporting structures, which are modeled internally as cladding, but it cannot be solved for the control rods, which had to be put in the bypass. For this reason, when evaluating the amounts of relocated materials in the core catcher design assumptions, it is always assumed that all the B4C from the control SAs relocated, even if it is not predicted in MELCOR.

MELCOR offers several possibilities of triggering the failure of the vessel lower head. Model used in this paper is triggering the failure when temperature inside of the lower head (mesh 3 of 8) reaches melting temperature of the RPV steel.

Because MELCOR does not allow other fissile material than $UO_2$ in the COR fuel components, physical properties of internal material $UO_2$ were changed to correspond with those valid for MOX with 25 % Pu content. More information on this particular issue can be found in Ref. 4.

Another important issue is connected to the recriticality of degraded core after the cladding of the fuel pins fails. The MELCOR model was set up with the total power in the core after reactor shutdown always equal to decay power only, recriticality issues are not solved within the scope of this study.

## IV. TRANSIENTS

In total, 4 cases of 3 types of transients were studied. They were expected to be the most penalizing transients in ALLEGRO, leading to severe core damage and RPV failure. They were selected without any attention paid to probability of such combinations of initiating and aggravating events. It is needed to emphasize here, that all the studied transients are far beyond the design basis of the reactor and some of them should even be excluded by design. They serve solely to study the ways and speed of propagation of the core degradation and to sett of proper conservative boundary conditions for the development of the core catcher.

The transients studied in this paper are:

a) LB-LOCA aggravated by SBO and failure of $N_2$ injection system

b) Water ingress (1 DHR loop) aggravated by SBO

c) Water ingress (2 DHR loops) aggravated by SBO

d) Failure of DHR check-valves opening aggravated by SBO

In case a), breach in one of the main cold legs is followed by SBO, i.e. unavailability of all active safety systems. Rundown of the main blowers keeps the main check-valves open for 45 seconds after reactor shutdown. then, cooling of the reactor is transferred from the main primary loops to the DHR loops in natural convection mode.

Cases b) and c) simulate loss of tightness in 1 or 2 He/water heat exchangers in the DHR system loops. 10 U-tubes in each exchanger is assumed to be damaged (resulting in 0,015 m2 of breach flow area). With no active systems available due to SBO and 1 or 2 DHR loops ineffective because of leaking water and isolation of the respective DHR secondary (water) circuits by check-valves, the reactor core is cooled by the remaining DHR loop(s) in natural convection mode after the main blowers rundown.

Case d) represents the boundary scenario with the fastest core degradation. Due to failure of opening of all the check-valves in the DHR loops, combined with SBO, the core is effectively isolated from any possible cooling loops - the main check-valves close automatically after the main blowers rundown (mass flow rate less than 3% of the nominal).

## V. RESULTS

Presented results are focused mainly on the timing and extent of core degradation. Results of analyses of cases a) - d) are described individually. The last subchapter summarizes amounts of materials relocated into the lower head of RPV after the support plate failure, which are used as an input for the analysis of core catcher options.

### V.A. LB-LOCA aggravated by SBO and $N_2$ injection failure

Depressurization of the reactor through the breach is very fast, pressures inside and outside the primary circuit are within 0,1 MPa after 13 seconds.
First melting of the cladding occurs 25 minutes after the reactor shutdown. The total extent of core degradation is shown in Fig. 4. The figure shows the nodes of COR package (core itself + the lower head of RPV), purple, yellow and orange color represents degraded material. It can be seen that the upper support plate does not fail despite degradation of the most of the core. The temperature of the supporting plate is decreasing from t = 15 h, reaching peak temperature 659 °C. The mechanical load is constant from t = 3,5 h. Temperatures of debris and melt in the degraded region reach maximum at t = 4,8 h.
From the above presented data, it can be reasonably expected that the degraded configuration shown in Fig. 4 (situation at t = 28 h) is the final one.

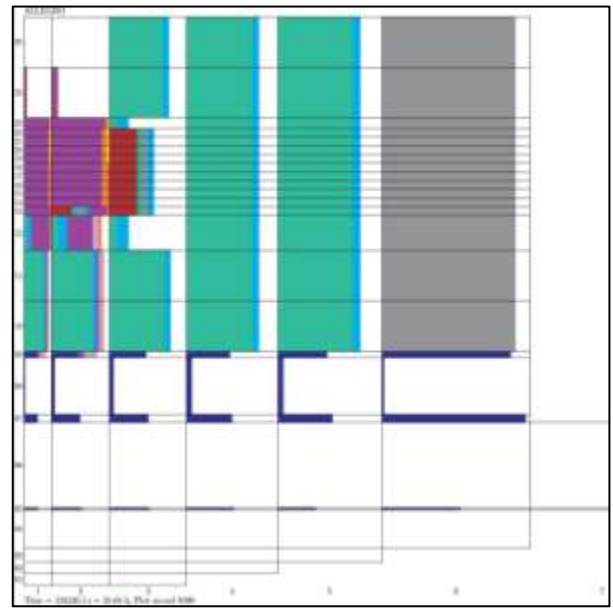

Fig. 4. Extent of the core degradation after 28h (LB-LOCA + SBO)

**V.B. Water ingress (1 DHR loop) aggravated by SBO**

This particular scenario has been already analyzed for the 75 MWth ALLEGRO[4]. The only change in the input, apart from the reduction of the reactor power, is different volume of water which cannot be isolated from the primary circuit. The new value is higher - 2,6 m3. It is due to an update in the design of the DHR water circuit. Now it is clear where the isolation valves in the circuit are located and the respective amount of water in the duct outside of the He/water HX has been added to the previously used value.

The progress of the accident is very similar to the calculations concerning the 75 MWth reactor. After the main blower rundown period, the core is cooled by 2 DHR loops in natural convection. The cooling capacity of the third loop is very limited because the respective water circuit is isolated 5 seconds after the break.

If the primary circuit is not depressurized, water, initially entering in the form of steam, starts to condensate in the cooler parts of the circuit and is being collected in the lowest part of the circuit, which is the lower head of the RPV. Large amounts of steam are then entering the core region and, as a consequence, steel cladding of the fuel starts to oxidize. This process causes gradual mechanical failure of the oxidized cladding, first occurring at $t = 1,5$ h. Progression of the scenario is quite slow, the upper support plate fails after 6 days. There was no failure of the bottom support plate in 7 days of the problem time calculated, but it can be reasonably expected, that the final state would be the same as in scenario c).

Depressurization of the primary circuit does not help the situation in any way, the final result and timing is approximately the same. On one hand, it prevents formation of the liquid water pool in the lower head and significantly reduces oxidation of the cladding and steel reactor internals, on the other hand, the overall cooling capacity of the DHR system in natural circulation dramatically decreases due to low helium density at lower pressure.

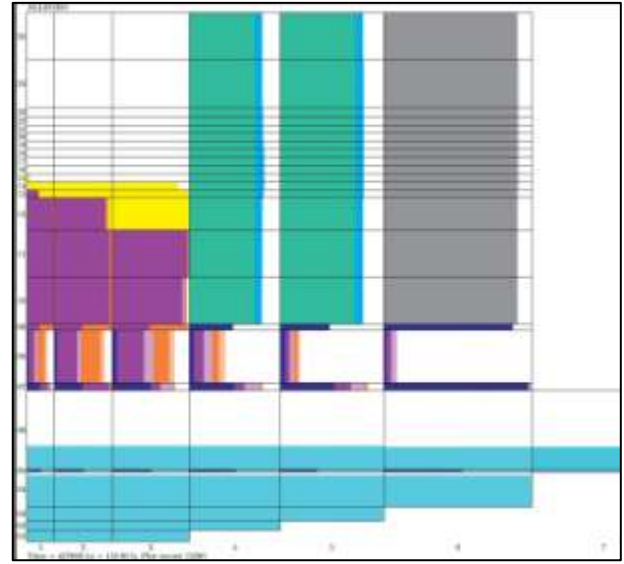

Fig. 5. Extent of the core degradation after 6 days (SBO + WI from 1 DHR loop)

**V.C. Water ingress (2 DHR loops) aggravated by SBO**

The scenario is very similar to the previous one, as well as the progression. The main qualitative difference is rather early failure of the supporting plates due to overtemperature. It happens mainly due to large amount of water in the lower head, surface of which is reaching the lower supporting plate and effectively blocking the flow of helium through the core. The only coolant flowing through the core from $t = 1900$ s is steam evaporated from the pool in the lower head.

The upper support plate fails at $t = 9$ h due to overtemperature, the lower one at $t = 14,5$ h due to mechanical overload. Debris and melt relocating into the lower head are quenched in the water pool, causing massive evaporation and further oxidation. The total decay heat is quite low after the pool is dried out ($t = 27$ h), namely 325 kW, so the lower head does not fail. The progress of core degradation is showed in Fig. 6, state showed in Fig. 6 d) can be considered as the final state.

| a) t = 9 h | b) t = 14,5 h | c) t = 16 h | d) t = 27 h |

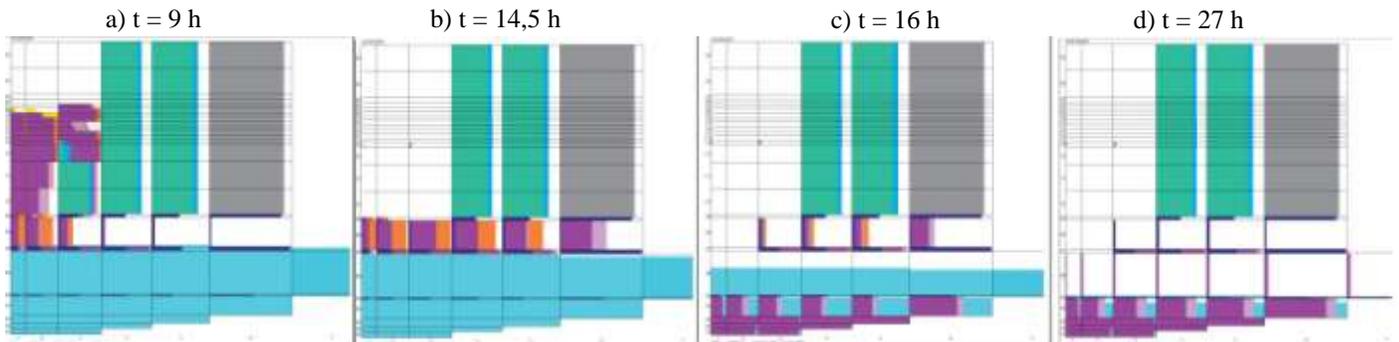

Fig. 6 a) to d) - Progression of the severe accident during scenario SBO + WI from 2 DHR loops

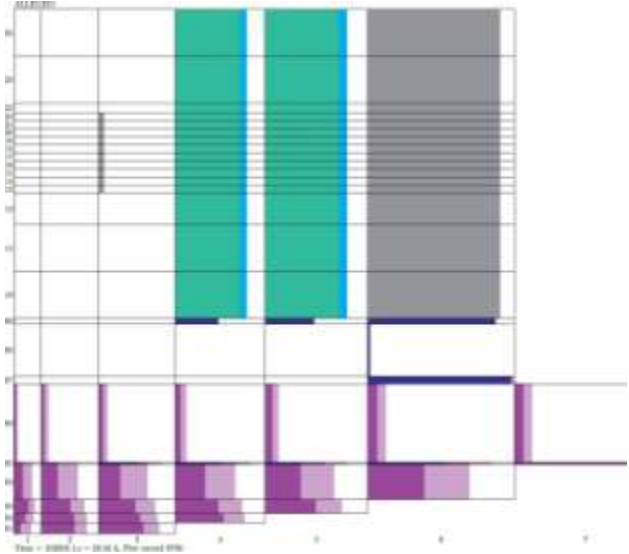

Fig. 7. Relocated core debris just before failure of the lower head (scenario failure of check-valves opening)

**V.D. Failure of DHR check-valves opening aggravated by SBO**

In this scenario, there are no means of reactor cooling after the first 45 s from SCRAM, apart from radiation from the primary circuit surface. Therefore, it should be the envelope scenario for the core catcher with the highest residual heat after the melt ejection, and highest demands on amount of the sacrificial material in the core catcher to prevent recriticality and to achieve a coolable composition of corium.

The propagation of the accident is very fast. Firt melting of the cladding occurs at t = 16,5 min., the upper support plate fails at t = 2,4 h due to overtemperature and the lower one at t = 8,5 h due to mechanical failure. The lower head fails at t = 18 h.

The effect of nitrogen injection is neglectable, its introduction results in postponing of lower head failure by approximately one hour. Fig. 7 shows the lower head after corium relocation, before the lower head failure.

**V.E. Materials relocated to the lower head**

TABLE 2 summarizes masses of materials (both melted and in form of debris) relocated to the lower head, either at the end of the calculation or at the time of ejection trough the vessel breach.

It can be seen that the total mass of the molten materials are very similar in both cases which lead to the failure of the support plates. It can be safely assumed from the data available for the other two cases, that the masses would be very similar if the two cases reached the relocation phase.

TABLE 2. Masses of materials relocated to the lower head.

| Material (kg) | case | | | |
|---|---|---|---|---|
| | a | b | c | d |
| **Fuel - solid** | - | - | 3137 | 3075 |
| **Fuel - melted** | - | - | 90 | 152 |
| **Steel - debris** | - | - | 12874 | 13222 |
| **Steel - melted** | - | - | 183 | 381 |
| **Steel oxide -solid** | - | - | 574 | 0 |
| **Steel oxide - melted** | - | - | 28 | 0 |
| **B4C** | - | - | 148 | 148 |
| **Total** | - | - | 17034 | 16830 |

**VI. CORE CATCHER PRE-DESIGN**

This chapter brings some considerations about the core catcher in ALLEGRO, supported by the results of analyses presented in chapter V. As an outcome, first draft of the possible core catcher design is presented.

**VI.A. Core catcher design basis**

A (definitely not complete) list of design basis criteria for the core catcher has been prepared and is available in TABLE 3.

Maximum CC dimensions are limited by both the dimensions of the GV and the position of the RPV inside the GV. A free space of 90 cm between the edge of the CC and the GV wall has been chosen as a safety margin. Vertical dimension of the CC is limited by the vertical position of the RPV and the CRDM located under it.

TABLE 3. Design basis criteria of the core catcher in ALLEGRO

| Criterion | Value |
|---|---|
| CC position | Inside GV, under RPV |
| Max. dimensions (height x radius) | 2 000 mm x 3 900 mm |
| Min. free volume | 3 m$^3$ |
| Heat removal capacity (at least) | 500 kW |
| Cooling system | Passive + indirect |

All the analyses done so far indicate, that the dependency of the amount of melted material reaching the lower head on the accident scenario is rather weak, and the total mass and volume of the material is around 17 000 kg and 2,5 m3 respectively. This value should be taken as the design basis, until a substantial change in the core design is made.

Case d) can be used to obtain the maximum decay power at the time of melt ejection to the CC. It is, conservatively, taken at the time of the lower support plate failure, assuming immediate lower head breach, and is equal to 420 kW.

As the free volume inside the GV is quite small, the indirect cooling of the corium in CC is preferable to avoid GV overpressurization by gasses released during direct corium cooling. The whole system should be fully passive to meet one of the GEN IV criteria: To achieve at least the safety level of the generation III reactors (which use passive core catcher cooling systems).

### VI.B. Core catcher pre-design assumptions

Evaluating the design basis criteria, the design of the core catcher wall cooling system similar to EPR seems to be an ideal option for the use in the ALLEGRO demonstrator. However, the integral concept of the EPR core catcher cannot be adopted here, because it requires a lot of free space under the RPV inside the GV.

Experiments described in Ref. 10 show that the passive part of the cooling system used in the EPR CC (channels inside the CC walls) is able to dissipate up to 80 kW/m$^2$ through the CC wall area. Presuming the maximal heat dissipation of 80 kW/m$^2$, the cooling area needed for a successful corium cooling during the minimal melting scenario is 6,25 m$^2$.

Based on the above-mentioned considerations, a pre-conceptual draft of the CC was proposed by UJV. The chosen shape is octangular. The sacrificial material forms a centered cone for better corium spreading across the CC surface. The sacrificial material has not been set so far. It is sure that it will have to dilute the oxidic part of the corium, which will compose mostly of the fuel debris/melt. Due to high risk of recriticality, it is also desirable to include neutron poison into the sacrificial material. One potential solution can be to use traditional combination of $Fe_2O_3$ and $Al_2O_3$, with addition of $B_2O_3$ as the neutron poison. Geometrical parameters can be found in TABLE 4, sketch of the CC in Fig. 8.

As can be seen from TABLE 4, the proposed CC meets the design basis criteria well. Area of the bottom of the CC is sufficient not only to dissipate the decay heat, but also to prevent a thick layer of corium. Typical corium composition coming from the scenarios analyzed in this paper will make 7 cm thick layer, allowing quite a lot of space for the sacrificial material.

TABLE 4. ALLEGRO core catcher draft dimensions

| | |
|---|---|
| Outer side lenght [m] | 2,8 |
| Total height [m] | 1,8 |
| Walls thickness [m] | 0,4 |
| Free volume (available for corium) [m$^3$] | up to 36,5* |
| Cooling area at the bottom of the CC [m$^2$] | 37,9 |

* (depends on the amount of sacrificial mat.)

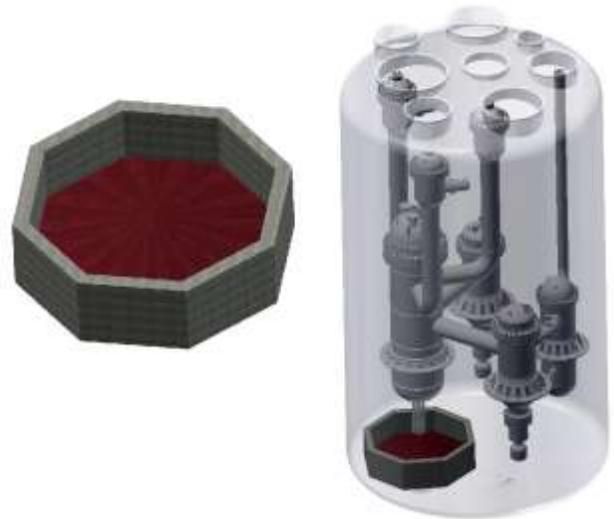

Fig. 8. Core catcher draft

### VII. DISSCUSION OF RESULTS

It can be seen from the results of same/similar transients, that the reduction of the thermal power (and power density) has certain positive effects, mainly in the depressurized transients, where the extent of the core degradation is significantly lower than in the 75 MWth version[4]. However, it has no practical impact on the results of the water ingress scenario. On the other hand, it has to be noted, that the water ingress with the DHR system in natural convection is, from the thermal-hydraulics point of view, very complicated one and the results of the presented analyses should be checked by other codes, preferably coupled with CFD, to simulate the system in more detail.

Another notable result is that most of the fuel remains solid throughout the accidents. It is due to quite a low thermal power of the reactor, resulting in small radial temperature gradient in the fuel pin after the reactor shutdown, and big difference between melting temperature of the cladding (1600 K) and the fuel (3000 K), with very limited chemical interactions.

It is very positive, that the only scenario leading to breach of the lower head and melt ejection, was the boundary "isolation" one - d)., which should be excluded by design. On the other hand, all the calculations proved the initial assumption, that the general way of degradation progress is downwards, with very little material melted in the radial direction, which is not preferable. It would be better to slow down the failure of the upper supporting plate to allow more time for melting of radial core internals such as the reflector subassemblies.

General mechanisms of RPV failure implemented in MELCOR are adapted to LWRs and are proven by vast amount of experiments. However, there are very little practical experience with severe accidents in reactors such as ALLEGRO. This issue will have to be addressed in the future.

Results of the accident calculations were used to define a set of design basis criteria of the core catcher. Previous conservative approach, dealing with two theoretical boundary melting scenarios, was replaced with best estimate approach building on the MELCOR calculations.

## VIII. CONCLUSIONS

Analysis of 4 cases of severe accidents in ALLEGRO with reduced power to 50 MWth using MELCOR 2.1 was presented. It focused on phenomenology of a severe accident in ALLEGRO with oxide core and stainless steel cladding, the computational part was focused on the timing and extent of the core degradation. It showed that the water ingress scenario is still a major problem even with the reduced power density. Results of LOCA scenarios are much more optimistic. Effects of elevated back-pressure in the guard vessel should be explored to further reduce the extent of core degradation during this type of accidents.

Results of the analysis were used as an input for considerations on core catcher in ALLEGRO. First draft of a possible solution of the core catcher was presented.

Presented work concerning the core catcher should be taken as the first step, more analyses of neutronic behavior, chemistry of the corium - sacrificial material interaction and detailed thermal-hydraulics will have to be done before the start of the experimental phase.

## ACKNOWLEDGMENTS

This work was financially supported by the project no.TE01020455 - CANUT, of the Technology Agency of the Czech Republic.